# LINEAR VLASOV SOLVER FOR MICROBUNCHING GAIN ESTIMATION WITH INCLUSION OF CSR, LSC AND LINAC GEOMETRIC IMPEDANCES

C. -Y. Tsai[#], Department of Physics, Virginia Tech, Blacksburg, VA 24061, USA
D. Douglas, R. Li, and C. Tennant, Jefferson Lab, Newport News, VA 23606, USA

*Abstract*

As is known, microbunching instability (MBI) has been one of the most challenging issues in designs of magnetic chicanes for short-wavelength free-electron lasers or linear colliders, as well as those of transport lines for recirculating or energy recovery linac machines. To more accurately quantify MBI in a single-pass system and for more complete analyses, we further extend and continue to increase the capabilities of our previously developed linear Vlasov solver [1] to incorporate more relevant impedance models into the code, including transient and steady-state free-space and/or shielding coherent synchrotron radiation (CSR) impedances, the longitudinal space charge (LSC) impedances, and the linac geometric impedances with extension of the existing formulation to include beam acceleration [2]. Then, we directly solve the linearized Vlasov equation numerically for microbunching gain amplification factor. In this study we apply this code to a beamline lattice of transport arc [3] following an upstream linac section. The resultant gain functions and spectra are presented here, and some results are compared with particle tracking simulation by ELEGANT [4]. We also discuss some underlying physics with inclusion of these collective effects and the limitation of the existing formulation. It is anticipated that this more thorough analysis can further improve the understanding of MBI mechanisms and shed light on how to suppress or compensate MBI effects in lattice designs.

## INTRODUCTION

The beam quality preservation is of a general concern in delivering a high-brightness beam through a transport line or recirculation arc in the design of modern accelerators. Microbunching instability (MBI) has been one of the most challenging issues associated with such beamline designs. Any source of beam performance limitations in such recirculation or transport arcs must be carefully examined in order to preserve the beam quality, such as the coherent synchrotron radiation (CSR), longitudinal space-charge (LSC) and/or other high-frequency impedances that can drive microbunching instabilities.

To accurately quantify the direct consequence of microbunching effect, i.e. the gain amplification factor,, we further extend our previously developed semi-analytical simulation code [1] to include more relevant impedance models, including CSR, LSC and linac geometric impedances. The LSC effect stems from (upstream) ripple on top of an electron beam and can accumulate an amount of energy modulation when the beam traverses a long section of a beamline. Such energy modulation can then convert to density modulation via momentum compaction $R_{56}$ downstream the beamline [5,6]. In addition, along the beamline, CSR due to electron radiation emission inside bending dipoles can have a significant effect on further amplifying such density to energy modulation [5,6]. The accumulation and conversion between density and energy modulations can possibly cause serious microbunching gain amplification (or, MBI).

___________________

[#] jcytsai@vt.edu

In this paper, we would first introduce the methods of microbunching gain calculation: first, a kinetic model based on linearized Vlasov equation [7,8] and, the other, particle tracking by ELEGANT as a benchmarking against our code. Then we briefly summarize the impedance models used in our simulations. After that, we illustrate simulation results, the gain functions and spectra for our example lattice, including a transport arc following a section of upstream linac. Finally, we discuss the underlying physics and summarize our observation from the simulation results.

## METHODS

To quantify the MBI in a transport or recirculation beamline, we estimate the microbunching amplification factor $G$ (or, bunching factor $g_k$) by two distinct methods. The first one, based on a kinetic model, is to solve a (linearized) Vlasov equation [7,8]. This method is of our primary focus in this paper. The second one, served as a benchmarking of the first method, is based on particle tracking method (here we use ELEGANT [4]). For the former, after mathematical simplification of the linearized Vlasov equation, we actually solve a general form of Volterra integral equation for the bunching factor. In our code, to facilitate us in simulating ERL-based lattices which usually contain vertical spreaders and/or recombiners, we extend the existing formulation to include both transverse horizontal and vertical bends. Also, we consider the presence of linac sections in a general beamline; the formulation of Volterra integral equation would be slightly modified [2] to accommodate RF acceleration or deceleration. . In sum, the governing equation for bunching factor $g_k$ is summarized below,

$$g_k(s) = g_k^{(0)}(s) + \int_0^s K(s,s')g_k(s')ds' \qquad (1)$$

where the kernel function can be particularly expressed as

$$K(s,s') = \frac{ik_0 C(s)}{\gamma_0} \frac{I_0 C(s')}{I_A} \hat{R}_{56}(s' \to s) Z(kC(s'),s') \times [\text{Landau damping}] \quad (2)$$

for the [Landau damping] term

$$\exp\left\{\frac{-k^2}{2}\left[\varepsilon_{x0}\left(\beta_{x0}\hat{R}_{51}^2(s,s') + \frac{\hat{R}_{52}^2(s,s')}{\beta_{x0}}\right) + \varepsilon_{y0}\left(\beta_{y0}\hat{R}_{53}^2(s,s') + \frac{\hat{R}_{54}^2(s,s')}{\beta_{y0}}\right) + \sigma_\delta^2 \hat{R}_{56}^2(s,s')\right]\right\} \quad (3)$$

with

$$\hat{R}_{56}(s' \to s) = \hat{R}_{55}(s')\hat{R}_{56}(s) - \hat{R}_{55}(s)\hat{R}_{56}(s') + \hat{R}_{51}(s')\hat{R}_{52}(s) - \hat{R}_{51}(s)\hat{R}_{52}(s') + \hat{R}_{53}(s')\hat{R}_{54}(s) - \hat{R}_{53}(s)\hat{R}_{54}(s') \quad (4)$$

$$\hat{R}_{5i}(s,s') = C(s)\hat{R}_{5i}(s) - C(s')\hat{R}_{5i}(s') \quad (5)$$

and the bunch compression factor

$$C(s) = \frac{1}{\hat{R}_{55}(s) - h_0 \hat{R}_{56}(s)} \quad (6)$$

Here the kernel function $K(s,s')$ describes relevant collective effects, $g_k(s)$ the resultant bunching factor as a function of the longitudinal position given a wavenumber $k$, and $g_k^{(0)}(s)$ is the bunching factor in the absence of collective effect (i.e. from pure optics effect). We particularly note that the above formulation can be applicable to the case with focusing in combined-function dipoles.

In the above formulation, we have made the coasting beam approximation, i.e. the modulation wavelength is assumed much shorter compared with the whole bunch duration. The transport functions $\hat{R}_{5i}(s)$ ($i = 1, 2, 3, 4, 6$) can be obtained directly from ELEGANT by tracking a set of (sufficient) number of macroparticles and deriving the 6 by 6 transport matrix at separate locations by proper transformation of the dynamic variables [2]:

$$\begin{bmatrix}\hat{x}\\\hat{y}\end{bmatrix} = \begin{bmatrix}x\\y\end{bmatrix}\sqrt{\frac{E_r(s)}{E_0}}; \begin{bmatrix}\hat{p}_x\\\hat{p}_y\end{bmatrix} \simeq \begin{bmatrix}p_x\\p_y\end{bmatrix}\sqrt{\frac{E_r(s)}{E_0}}; \hat{z} = z; \hat{\delta} = (\delta+1) - \frac{E_r(s)}{E_0} \quad (7)$$

Here $E_0$ is the initial beam energy and $E_r(s)$ is the reference energy at a specific location $s$. $\delta \equiv (E - E_0)/E_0$. Note here that the transformation assumes the energy gain due to acceleration (or deceleration) varies slowly, i.e. $1/E_r \, dE_r/ds \ll 1$.

To quantify MBI in a single-pass system, we define the microbunching gain as functions of the global longitudinal coordinate s as well as the initial modulation wavelength $\lambda$ (or, $k = 2\pi/\lambda$):

$$G(s, k = 2\pi/\lambda) \equiv \left|\frac{g_k(s)}{g_k^{(0)}(0)}\right| \quad (8)$$

Hereafter, we simply call $G(s)$ the gain function, which is a function of s for a given modulation wavenumber, and denote $G_f(\lambda)$ as the gain spectrum, which is a function of modulation wavelength at a specific location (e.g. denoted with a subscript "$f$" at the exit of a beamline). It is worth to mention the general physical meaning of Eqs. (1-3): a density perturbation at $s'$ induced an energy modulation through a collective impedance $[Z(k(s'))]$ and is subsequently converted into a further density modulation at $s$ via non-vanishing momentum compaction $R_{56}(s' \to s)$.

## IMPEDANCE MODELS

For an electron beam traversing an individual dipole, CSR can have both steady-state and transient effects. In addition, when a beam goes through a long transport line, LSC can have a significant effect on accumulating energy modulations. Moreover, when a beam is accelerated the RF cavity is characteristic of the (high-frequency) geometric impedance which can also accumulate an amount of energy modulation. Here we quote the relevant impedance expressions without further derivation:

### CSR in Free Space

For a relativistic electron beam ($\beta = 1$, $\gamma < \infty$) traversing a bending dipole, the free-space steady-state CSR impedance per unit length can be expressed as [9]:

$$\text{Re}\left[Z_{CSR}^{s.s.NUR}(k(s);s)\right] = \frac{-2\pi k(s)^{1/3}}{|\rho(s)|^{2/3}} \text{Ai}'\left(\frac{(k(s)|\rho(s)|)^{2/3}}{\gamma^2}\right) + \frac{k(s)\pi}{\gamma^2}\left(\int_0^{(k(s)|\rho(s)|)^{2/3}/\gamma^2} \text{Ai}(\varsigma)d\varsigma - \frac{1}{3}\right) \quad (9)$$

$$\text{Im}[Z_{CSR}^{s.s.NUR}(k(s);s)] \simeq \frac{2\pi k(s)^{1/3}}{|\rho(s)|^{2/3}}\left\{\frac{1}{3}\text{Bi}'(x) + \int_0^x\begin{bmatrix}\text{Ai}'(x)\text{Bi}(t)\\-\text{Ai}(t)\text{Bi}'(x)\end{bmatrix}dt\right\}$$

where $x = (k(s)|\rho(s)|)^{2/3}/\gamma^2$, $k = 2\pi/\lambda$ is the modulation wavenumber, $\rho(s)$ is the bending radius, and Ai and Bi are Airy functions. Under ultrarelativistic approximation ($\gamma \to \infty$), Eq. (9) is reduced to the well-known expression [10,11]

$$Z_{CSR}^{s.s.UR}(k(s);s) = \frac{-ik(s)^{1/3}A}{|\rho(s)|^{2/3}}, A = -2\pi[\text{Bi}'(0)/3 + i\text{Ai}'(0)] \quad (10)$$

Prior to reaching steady state, the beam entering a bend from a straight section would experience the so-called entrance transient state, where the impedance per unit length can be obtained by Laplace transformation of the corresponding wakefield [12-14]:

$$Z_{CSR}^{ent}(k(s);s) = \frac{-4}{s^*}e^{-4i\mu(s)} + \frac{4}{3s^*}(i\mu(s))^{1/3}\Gamma\left(\frac{-1}{3}, i\mu(s)\right) \quad (11)$$

where $\mu(s) = k(s)z_L(s)$, $s^*$ is the longitudinal coordinate measured from dipole entrance, $z_L = (s^*)^3/24\rho^2$, and $\Gamma$ is the upper incomplete Gamma function.

There are also exit CSR transient effects as a beam exits from a dipole. For the case with fields generated from an upstream electron (at retarded time) propagating across the dipole to downstream straight section, i.e. Case C of Ref. [14], the corresponding impedance per unit length can be similarly obtained by Laplace transformation:

$$Z_{CSR}^{exit}(k(s);s) = \frac{-4}{L_b + 2s^*}e^{\frac{-ik(s)L_b^2}{6|\rho(s)|^2}(L_b + 3s^*)} \quad (12)$$

where $s^*$ is the longitudinal coordinate measured from dipole exit and $L_b$ is the dipole length.

For the impedance expression of the case where fields generated from an electron (at retarded time) within a dipole propagating downstream the straight section, we use the following expression for the exit transient impedance [15]:

$$Z_{CSR}^{drif}(k(s);s) \approx \begin{cases} \frac{2}{s^*}, & \text{if } \rho^{2/3}\lambda^{1/3} \leq s^* \leq \lambda\gamma^2/2\pi \\ \frac{2k(s)}{\gamma^2}, & \text{if } s^* \geq \lambda\gamma^2/2\pi \\ 0, & \text{if } s^* < \rho^{2/3}\lambda^{1/3} \end{cases} \quad (13)$$

where $s^*$ is again the longitudinal coordinate measured from dipole exit. This expression assumes the exit impedance comes primarily from coherent edge radiation in the near-field region (i.e. $z < \lambda\gamma^2$), and in our simulation we only include transient effects right after a nearby upstream bend.

Here we note that these CSR models are valid only when the wall shielding effect is negligible. The wall shielding effect becomes important when the distance from the beam orbit to the walls $h$ satisfies $h \leq (\rho\lambda^2)^{1/3}$. In this situation, one should consider to use the shielded CSR impedance in evaluating the CSR-induced microbunching gain. Currently we adopt the steady-state impedance based on parallel-plate model [16,17] as

$$Z_{CSR}^{p.p.}(k) = \frac{8\pi^2}{h}\left(\frac{2}{k(s)\rho(s)}\right)^{\frac{1}{3}}\sum_{p=0}^{\infty}F_0(\beta_p), \ \beta_p = (2p+1)\frac{\pi}{h}\left(\frac{\rho(s)}{2k^2(s)}\right)^{\frac{1}{3}} \quad (14)$$

where

$$F_0(\beta) = \text{Ai}'(\beta^2)\left[\text{Ai}'(\beta^2) - i\text{Bi}'(\beta^2)\right] + \beta^2\text{Ai}'(\beta^2)\left[\text{Ai}(\beta^2) - i\text{Bi}(\beta^2)\right].$$

For a more realistic beam pipe geometry, such as rectangular cross section, we consider to load impedance data externally calculated by other dedicated program. We are also working in progress on a numerical approach to obtain the impedance model which takes into account the straight-bend-straight section with rectangular beam pipe [18]. The results shall be valid for non-ultrarelativistic and finite wall resistivity, including both CSR and LSC.

*LSC in Free Space*

Below we present two slightly different LSC impedance expressions [19] implemented in our code. The first one is on-axis model, which assumes a transversely uniform density with circular cross section of radius $r_b$,

$$Z_{LSC}^{on-axis}(k(s);s) = \frac{4i}{\gamma r_b(s)}\frac{1 - \xi K_1(\xi)}{\xi} \quad (15)$$

where $\xi = \frac{k(s)r_b(s)}{\gamma}$ and $r_b(s) \approx \frac{1.747}{2}(\sigma_x(s) + \sigma_y(s))$ [20]. The second one is the average model, which integrates the radial dependence [19,21],

$$Z_{LSC}^{ave}(k(s);s) = \frac{4i}{\gamma r_b(s)}\frac{1 - 2I_1(\xi)K_1(\xi)}{\xi} \quad (16)$$

where $\xi$ is defined above the same way. In the following we use the first model, in accordance with ELEGANT.

*Linac Geometric Effect*

If a beam experiences acceleration, deceleration or chirping along a section of a linac with RF cavities, the periodic structure in general features a geometric impedance. We consider this collective effect by adopting the following expression [22-24],

$$Z_{linac}^{UR}(k) = \frac{4i}{ka^2}\frac{1}{1+(1+i)\frac{\alpha L}{a}\sqrt{\frac{\pi}{kg}}} \quad (17)$$

where $\alpha \approx 1 - 0.4648\sqrt{g/L} - 0.07 g/L$, $a$ is the average iris radius, $g$ is the gap distance between irises, and $L$ is the cell/period length.

## SIMULATION RESULTS

In this section we take a high-energy transport arc [3] following an upstream linac section as an example for our analysis. Table 1 summarizes some beam parameters used in our simulations. In this beamline lattice, the electron beam is accelerated from 50 MeV to 1.11 GeV throguh a section of linac including 200 cavities with assumed voltage gradient ~10MV/m, frequency 1497 MHz and on-crest acceleration phase. We note that, at the exit of the linac, beam phase-space distribution is not particularly optimized to match that of the downstream transport arc. However, the purpose here is to demonstrate the microbunching gain evolution along a general beamline with inclusion of beam acceleration. Thus, this imperfect match should not be a concern. The transport arc is a second-order achromat and being globally isochronous with a large dispersion modulation across the entire arc. Figure 1 shows the dispersion $R_{16}(s)$ and momentum compaction functions $R_{56}(s)$ along the linac-arc beamline. It can be seen in Fig. 1 that the transport function $R_{56}(s)$ has taken non-ultrarelativistic effect into consideration.

Table 1: Initial beam parameters for the example lattice

| Name | Value | Unit |
|---|---|---|
| Beam energy (at linac entrance) | 50 | MeV |
| Beam energy (at linac exit) | 1.11 | GeV |
| Peak bunch current | 88 | A |
| Normalized emittance | 0.3 | μm |
| Initial beta function | 18 | m |
| Initial alpha function | -3.6 | |
| Energy spread (uncorrelated) | ~3×10$^{-4}$ | |

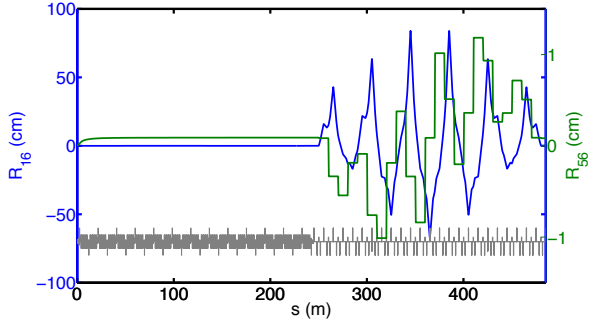

Figure 1: Dispersion (blue) and momentum compaction (green) functions of the example linac-arc lattice.

Figures 2 and 3 show the evolution of gain functions $G(s)$ along the beamline. ELEGANT tracking simulation was performed for a Gaussian beam of 70M macroparticles with flattop z-distribution. LSC effect is only considered within drift elements and RF cavities. In other elements such as dipoles, focusing or defocusing quads this effect is neglected. In Fig. 3, the CSR effects include both transient and steady states inside individual dipoles, as well as exit transient effect in the downstream drift sections. We found in Fig.2 the gain is slightly decreased in the linac section, because of LSC-induced plasma oscillation along with beam acceleration (see also $R_{56}$ in Fig. 1). Our Vlasov solutions match well with ELEGANT tracking results throughout the lattice except some "special" locations (see, e.g. at s = 410-440 m, in Figs. 2 and 3). After carefully examining numerical parameters to ensure the convergence of ELEGANT tracking results, we found the gain deviation (between our Vlasov solutions and ELEGANT) does not stem from numerical issues but is due to physical non-uniformity of the bunch profile as a result of RF curvature. Here we assume this RF curvature from the linac is not compensated by a harmonic cavity, as is done in several linac-based FELs.

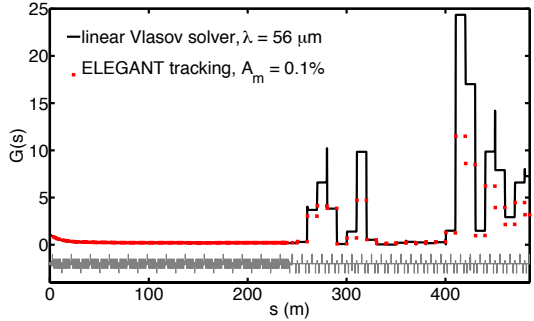

Figure 2: LSC-induced microbunching gain function $G(s)$ for the linac-arc lattice.

Here we would show that this bulk non-uniformity can cause the microbunching gain a bit reduced, compared with that of our Vlasov solution which does exclude this effect from gain estimation. Due to the presence of RF cavity, the accelerated beam is characteristic of an RF curvature (see top figure of Fig. 4). In our case with on-crest acceleration, we can simply assume the particle energy deviation related to its longitudinal coordinate to be

$$\delta_i = hz_i + qz_i^2 \qquad (18)$$

where the linear chirp $h$ vanishes but quadratic chirp $q$ does exist (< 0 in our case). With this ($z$-$\delta$) correlation, we can define an effective (and local) chirp to be

$$h^{eff}(z_i) \equiv -\frac{\partial \delta_i}{\partial z_i} = -2qz_i \Rightarrow \begin{cases} < 0, \text{ for bunch tail } (z_i < 0) \\ > 0, \text{ for bunch head } (z_i > 0) \end{cases} \qquad (19)$$

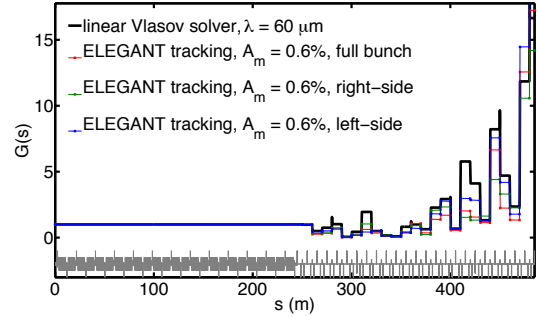

Figure 3: CSR-induced microbunching gain function G(s) for the linac-arc lattice. Here the CSR effects include both entrance and exit transients as well as steady-state effect.

The local bunch compression factor can be described as

$$C(s,z_i) = \frac{1}{R_{55}(s) - h^{eff}(z_i)R_{56}(s)} \Rightarrow \begin{cases} > 1, \text{ for bunch head \& } R_{56}(s) > 0 \\ < 1, \text{ for bunch tail \& } R_{56}(s) > 0 \end{cases} \qquad (20)$$

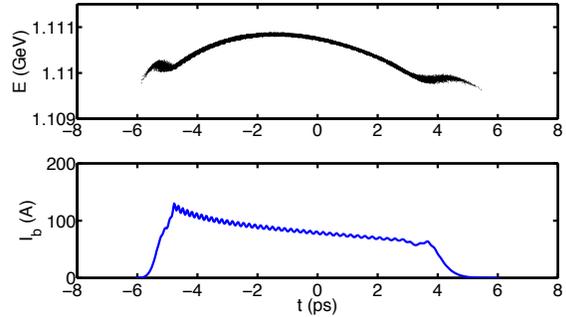

Figure 4: (Top) longitudinal phase space distribution at s = 410 m. (Bottom) bunch current density. Note here the bunch head is to the left.

By the above simple analysis we can explain the presence of bunch non-uniformity in Fig. 4. Note that in Fig. 4 the bunch head is to the left since we use $t$ as the variable instead of $z$. Now we would like to illustrate how such bulk non-uniformity affects the bunch spectrum (or, the bunching factor) and thus the microbunching gain.

Without resorting to rigorous mathematical derivation, we only illustrate numerically the difference of bunch spectra between uniform density-modulated and non-uniform density-modulated bunch profiles. Figure 5 highlights the following three different functions:

$$f_1(t) = A_0 + A_1 \sin(\omega t), \quad f_2(t) = A_0 t + A_1 \sin\left[(\omega - A_2 t)t\right] \qquad (21)$$
$$f_3(t) = A_0 t + A_1 \sin\left[(\omega + A_2 t)t\right]$$

where $f_1$ describes the coasting-beam model, while $f_2$ and $f_3$ approximately represent the non-uniformity of the bunch profile (in bottom figure of Fig. 4, corresponding to the right-side and left-side of the bunch profile, respectively).

From Fig. 5 it can be obviously seen that with non-uniform bunch profile ($f_2$ and $f_3$) the corresponding bunch spectral amplitudes become always smaller than the uniform one. This indeed causes the microbunching gain a bit reduced at certain locations with local bunch compression.

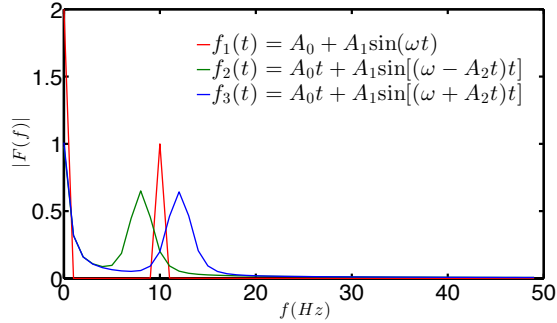

Figure 5: Bunch spectra (as Fourier transformation) of three different functions in Eq. (21). Assume $A_0=A_1=1$, $A_2=4\pi$, and the nominal frequency $f = 10$ Hz.

Figures 6-8 show the microbunching gain spectra including different collective effects. In Fig. 6, we can see the dependence of modulation wavelength on LSC-induced microbunching gain. In Fig. 7, both our Vlasov solver and ELEGANT include all relevant CSR effects, i.e. entrance, exit transient, and steady-state CSR effects. We believe the difference comes from: (i) non-uniformity of the bulk bunch; (ii) the different models of exit transient CSR used in our simulation [see Eq. (13)] and in ELEGANT [4, 14]. We also notice the gain reduction of non-ultrarelativistic (NUR, black curve) CSR compared with ultrarelativistic (UR, blue curve) approximation. The fluctuations shown in Fig. 7 are likely due to the CSR drift model we applied [Eq. (13)], which requires further investigation. Thus far in ELEGANT, it is not trivial to include all relevant collective effects such as CSR, LSC and linac geometric wakes into thorough consideration for MBI estimation. However, with our linear Vlasov solver, it is straightforward to add these relevant impedance models into consideration. In Fig. 8 we add these collective effects altogether for microbunching gain estimation. We observe that the overall microbunching gain is in fact an accumulation effect of density-energy conversion throughout the beamline. In the long section of the upstream linac, LSC and linac geometric wake have accumulated an amount of energy modulation, and later such energy modulation converts to density modulation through the downstream nonvanishing $R_{56}$ [e.g. at s = 280 m in Fig. 2]. Then, the modulation can be further amplified via CSR effect (in this case, mainly steady-state CSR) downstream the bends. Note that with the large gain shown in Fig. 8 the microbunching may reach nonlinear regime where linearized Vlasov solution is no longer valid for a practical point of view. For the validity of linear analysis, we assume the initial perturbation is sufficiently small (although in fact it may not be so small) that the magnitude of the bunching factor along the beamline does not exceed a certain value.

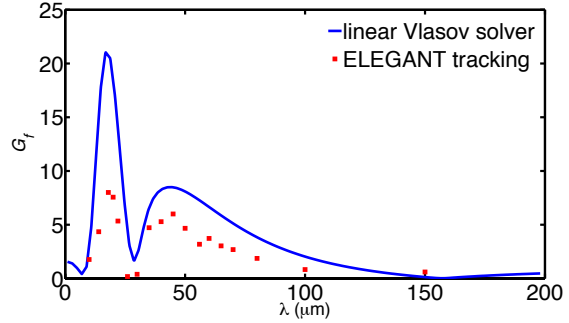

Figure 6: Microbunching gain spectra with LSC effects. Note here that in ELEGANT simulation we vary the initial modulation amplitudes around 0.1-0.6%.

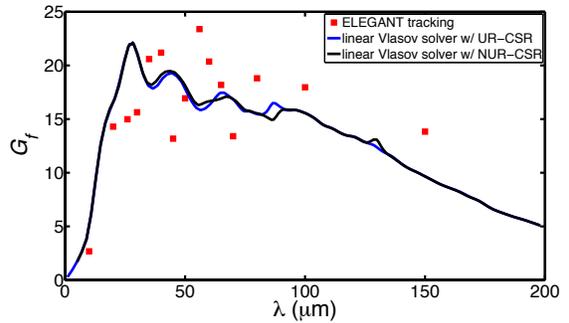

Figure 7: Microbunching gain spectra with all relevant CSR effects. ELEGANT results include both entrance and exit transient as well as steady-state impedances. The initial modulation amplitudes are varied around 0.1-0.6% to ensure numerical convergence.

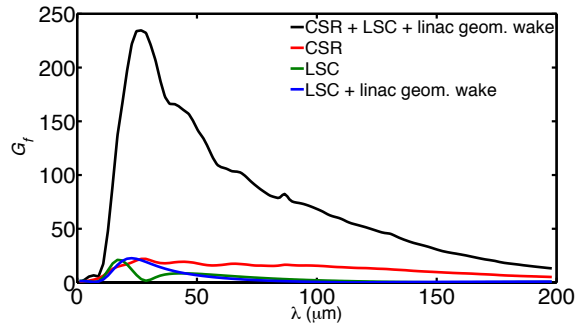

Figure 8: Microbunching gain spectra with various combinations of collective effects. To simulate the gain with linac geometric impedance, here we assume the linac parameters are: $a$ = 3.07 cm; $L$ = 10.0 cm; $g$ = 8.0 cm; $\alpha$ = 0.528.

## SUMMARY AND CONCLUSION

We have presented the theoretical formulation to include the case with beam acceleration based on the scaled dynamical variables [2]. Then, we summarized various collective impedance models relevant to the microbunching instability for our subsequent analysis. After that, we demonstrated a linac-arc beamline as our microbunching gain estimation. Our simulation results match well with ELEGANT tracking except at some locations with local bunch compression. We identify that such local bunch compression is due to non-uniformity of the bunch profile which stems from RF curvature in the upstream acceleration section, and can result in microbunching gain reduction. With inclusion of all relevant collective effects (CSR, LSC and linac geometric effects) in the simulation, the results show that microbunching gain can be significantly enhanced due to energy-density conversion along the beamline.

## ACKNOWLEDGMENT

We thank Steve Benson for bringing many motivating issues for our studies. This work is supported by Jefferson Science Associates, LLC under U.S. DOE Contract No. DE- AC05-06OR23177.